 \documentclass[prb,floatfix,preprint,nofootinbib,12pt]{revtex4}

\usepackage{graphicx}
\usepackage{epstopdf}
\usepackage{amsmath}
\usepackage{enumitem}
\usepackage{datetime}

\begin{document}

\newcommand{\etal}{{\it et al.}\/}
\newcommand{\gtwid}{\mathrel{\raise.3ex\hbox{$>$\kern-.75em\lower1ex\hbox{$\sim$}}}}
\newcommand{\ltwid}{\mathrel{\raise.3ex\hbox{$<$\kern-.75em\lower1ex\hbox{$\sim$}}}}
\renewcommand{\thefootnote}{\fnsymbol{footnote}}

\title{ \bf Simplicial  Quantum Gravity\protect\footnote{Talk delivered at the 3rd Moscow Quantum Gravity Seminar, October 25, 1984.}}

\author{James B. Hartle}
\affiliation{Department of Physics, University of California, Santa Barbara, CA 93106-9530, USA}

\begin{abstract}
Simplicial approximation and the ideas associated with the Regge calculus Refs(3,11,12)
provide a concrete way of implementing a sum-over-histories formulation of 
quantum gravity. A simplicial geometry is made up of flat simplices joined
together in a prescribed way together with an assignment of lengths to their
edges. A sum over  simplicial geometries is a sum over the different ways the
simplices can be joined together with an integral over their edge lengths.
The construction of the simplicial Euclidean action for this approach to
quantum general relativity is illustrated. The recovery of the diffeomorphism
group in the continuum limit is discussed. Some possible classes of simplicial
complexes with which to define a sum over topologies are described. In two
dimensional quantum gravity it is argued that a reasonable class is the class of pseudomanifolds.

Note Added:   {\it Provenance:  This paper is essentially a historical document. As mentioned in the footnote, this article started out as a conference talk well before arXiv. It is closely connected to the author's paper Ref(14).  It appears in print here  for the first time nearly 40 years later for two  reasons:  First it is a simple, short, but still current, exposition  of the use of lattice gravity  to sum over topology as discussed for example in the author's``My Time Line in Quantum Mechanics''  Ref(9).  The second reason is that  Figures 2  and 3 show numerical calculations  of the action for different simplicial geometries calculated by the author  but not otherwise readily available. 
A third reason is that there seems to be renewed interest in summing over topologies among those working in quantum gravity and quantum cosmology.   The article is unchanged from the original text except to update some terminology,  to cite some newer more relevant references,  and to divide the exposition into two more manageable parts.  The first part deals with sums over geometries, the second part with sums over topologies e.g.  Ref(14) .  No attempt had been made to cite later work that bears on the questions raised here e.g. simplicial conifolds Ref(13). }\end{abstract}


\maketitle



\section{Summing over Geometries} 

The sum over histories formulation of quantum mechanics provides a direct and
general framework for the construction of a quantum theory of gravity. Quantum
amplitudes are specified by sums over geometries in a class appropriate to the
particular amplitude of interest. For example, the amplitude for a given three
geometry $^{(3)}{\cal G}$ to occur in the state of minimum excitation of a closed
cosmology is Eq(1) and Ref(1).
\begin{equation}
  \psi\left[\!^{(3)}{\cal G}\right]=\sum_{\cal G}\exp\left(-I[{\cal G}]\right).
	\label{eq:1}
\end{equation}
Here, $I$ is the Euclidean gravitational action and the sum is over all
connected, compact Euclidean four geometries $\cal G$ which have the given
three-geometry as a boundary. This is the no-boundary  wave
function of our Universe.  Ref(1). 

A four-geometry is a four-dimensional manifold with a metric. A sum over
geometries therefore means a sum over four-manifolds and a functional integral
over physically distinct four-metrics. To understand what such sums and
integrals mean, one should have a practical method of implementing them.
Simplicial approximation and the ideas associated with the Regge
calculcus Refs(3,11,12) and Ref(7)   provide such a method.  I would like to
illustrate their utility. I shall emphasize the use of simplicial methods as
tools for definition, approximation, and  calculation in a continuum theory of
gravity. It may be, however, that the discrete version is more fundamental
and the continuum only an approximation as, for example, in a theory with a
fundamental length.

A simplicial geometry is made up of flat simplices joined together. A two
dimensional surface can be made out of flat triangles. A three-dimensional
manifold can be built out of tetrahedra; in four dimensions one uses
4-simplices and so on. The information about topology is contined in the rules
by which the simplices are joined together. A metric is provided by an
assignment of edge lengths to the simplices and a flat metric to their
interiors. With this information we can, for example, calculate the distance
along any curve threading the simplices.

\begin{figure}[htbp]
  \includegraphics[width=7.5cm]{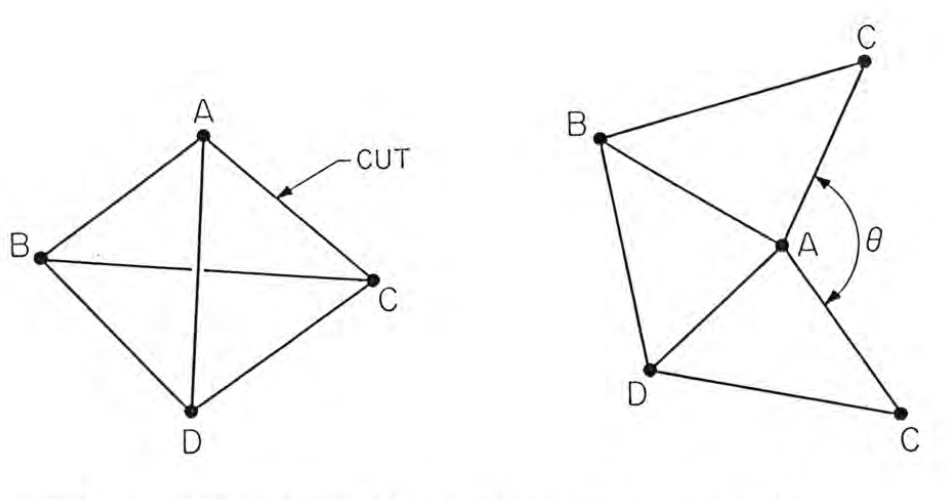}
  \caption{The surface of a tetrahedron is a two-dimensional surface whose
	curvature is concentrated at its vertices. To flatten the three triangles
	meeting at vertex A one could cut the tetrahedron along edge AC. The angle
	$\theta$ by which the edges AC fail to meet when flattened is a measure of
	the curvature at A called the deficit angle.
	\label{fig:1}}
\end{figure}

A two-dimensional surface made up of triangles is curved  in general  as, for
example, the surface of the tetrahedron in Figure~\ref{fig:1}. 
The curvature is not in the interior of the triangles; they are flat. It is not
on the edges; two triangles meeting in a common edge can be flattened without
distorting them. Rather, the curvature of a two-dimensional simplicial geometry
is concentrated at its vertices, because one cannot flatten the triangles
meeting in a vertex without cutting one of the edges. If one does cut one of
the edges and flatten then the angle by which the separated edges fail to meet
is a measure of the curvature called the deficit angle. (See Figure~\ref{fig:1}.)
It is the angle by which a vector would be rotated if parallel transported
around the vertex. Concretely the deficit angle is $2\pi$ minus the sum of the
interior angles of the triangles meeting at the vertex. It can thus be expressed
as a function of their edge lengths.

In four dimensions the situation is similar with all dimensions increased by 2.
The geometry is built from flat 4-simplices. Curvature is concentrated on the
two-dimensional {\it triangles}\/ in which they intersect. There is a deficit
angle associated with each triangle which is $2\pi$ minus the sum of the
interior angles between the bounding tetrahedra of the 4-simplices which
intersect the trangle.

The gravitational action may be expressed as a function of the deficit angles
and the volumes of the simplices. For example, the Euclidean Einstein action
with comological constant for a connected closed manifold in $n$\/-dimensions is,
\begin{equation}
  g_n\ell^{n-2}_pI_n=-\int d^nx(g)^{1/2}(R-2\Lambda).\label{eq:2}
\end{equation}
Here, $\ell_p=(16\pi G)^{1/2}$ is the Planck length and $g_n$ is a dimensionless
coupling. On a simplicial geometry  eq(2) becomes exactly Ref(3). 
\begin{equation}
  g_n\ell^{n-2}_pI_n=-2 \sum_{\sigma\epsilon\sum_{n-2}}V_{n-2}\theta_{n-2}+
	2\Lambda\sum_{\tau\epsilon\sum_n}V_n.\label{eq:3}
\end{equation}
Here, $\sum_k$ is the collection of $k$-simplices and $V_k$ is the volume of a
$k$-simplex. The deficit angle $\theta_k$ is defined by
\begin{equation}
  \theta_k(\sigma)=2\pi-\sum_{\tau\supset\sigma}\theta_k(\sigma,\tau), Eq(3).
\end{equation}
where the sum is over all the $(k+2)$-simplices $\tau$ which meet $\sigma$ and the
$\theta_k(\sigma,\tau)$ are their interior angles at $\sigma$. Both $V_k$ and
$\theta_k(\sigma,\tau)$ are simply expressible in terms of the edge lengths
through standard flat space formulae. By using these expressions in Eq(3)
the action becomes a function of the edge lengths. Other gravitational actions,
such as curvature squared Lagrangians, may be similarly expressed---not exactly
as here, but in an approximate form which becomes exact in the continuum limit, as in Regge's original paper, Ref(3). .

Sums over geometries may be given concrete meaning by taking limits of sums of
simplicial approximations to them. This is analogous to defining the Riemann
integral of a function as the limit of sums of the area under piecewise linear
approximations to it. Consider, by way of example, the sum over four geometries
which gives the expectation value of a physical quantity $A[{\cal G}]$ in the state
of minium excitation for closed cosmologies Ref(1).
\begin{equation}
    \langle A\rangle=\frac{\sum_{\cal G}A[{\cal G}]\exp(-I[{\cal G}])}
		                      {\sum_{\cal G}\exp(-I[{\cal G}])},\label{eq:5}
\end{equation}
where the sum is over compact, closed Euclidean four-geometries. We are accustomed
to think of a geometry as a manifold with a metric, and one might therefore want
to think of the sum in (\ref{eq:5}) as a sum over closed manifolds and a sum over
physically distinct metrics on those manifolds. Simplicial approximation could be
used to give a concrete meaning to such a sum as follows: (1) Fix a number of
vertices $n_0$. (2) Approximate the sum over manifolds $M$ as the sum over the number
of way of putting together 4-simplices so as to make a simplicial manifold with
$n_0$ vertices. (3) Approximate the sum over physically distinct metrics by a
multiple integral over the squared edge lengths $s_i$. (4) Take the limit of these
sums as $n_0$ goes to infinity. In short, express $(A)$ as
\begin{equation}
  	\langle A\rangle=\lim_{n_0\to\infty}
		\frac{\sum_{M(n_0)}\int_Cd\sum_1A(s_i,M)\exp[-I(s_i,M)]}
		{\sum_{M(n_0)}\int_Cd\sum_1\exp[-I(s_i,M)]}\label{eq:6}
\end{equation}
There remains the specification of the measure and the contour
$C$ for the integral over edge lengths. Of course, today we understand little
about the convergence of such a process but it is at least definite enough to be
discussed.

The central ingredient in weighting the sum over geometries is the action. A
variety of gravitational actions could be considered which correspond in the
continuum limit to Einstein's action or curvature squared Lagrangians and more
complicated actions. The extrema of the action are the solutions of a finite
set of algebraic equations
\begin{equation}
   \frac{\partial I}{\partial s_i}=0.\label{eq:7}
\end{equation}
For the Regge action (\ref{eq:3}) in four dimensions these are the discrete
version of the Einstein's equation
\begin{equation}
   \sum_{\sigma\epsilon\sum_4}\theta(\sigma)\frac{\partial V_2}{\partial s_i}=
	 \lambda\sum_{\tau\epsilon\sum_4}\frac{\partial V_4}{\partial s_i}.\label{eq:8}
\end{equation}
These extrema can be used to construct the semiclassical approximation to the
quantum theory. (For some later efforts at solving the equations see, e.g. 
Ref(8)).

\begin{figure}[htbp]
  \includegraphics[width=7.5cm]{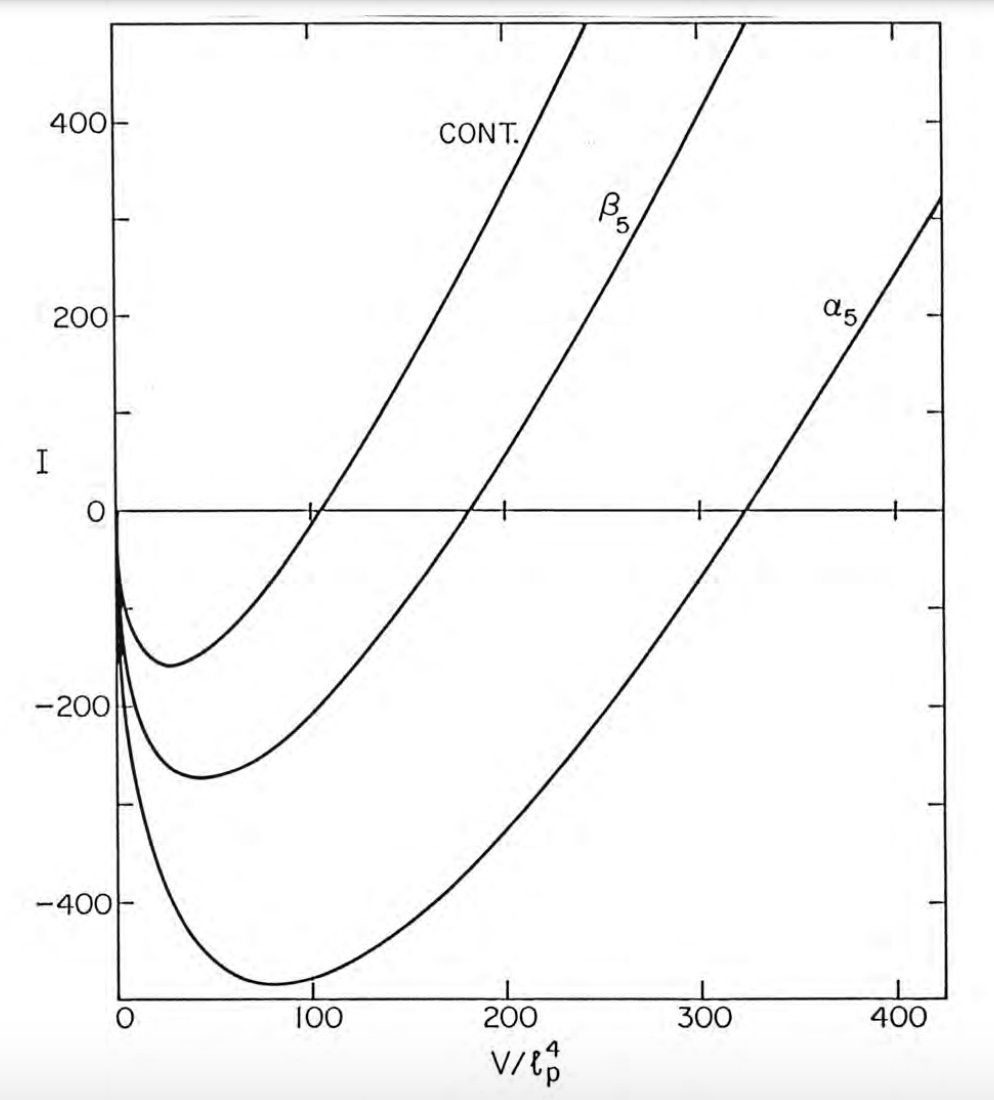}
  \caption{The action for some homogeneous isotropic four geometries as a
	function of volume. The figure shows the action for the 4-geometries which
	are the boundary of a 5-simplex  denoted  by ($\alpha_5$) and the 5-dimensional cross
	polytope ($\beta_5$) (the 5-dimensional generalization of the octohedron)
	when all of their edges are equal. The ``continuum" action for
	the 4-sphere is also plotted. 
	\label{fig:2}}
\end{figure}

\begin{figure}[htbp]
  \includegraphics[width=5.5in]{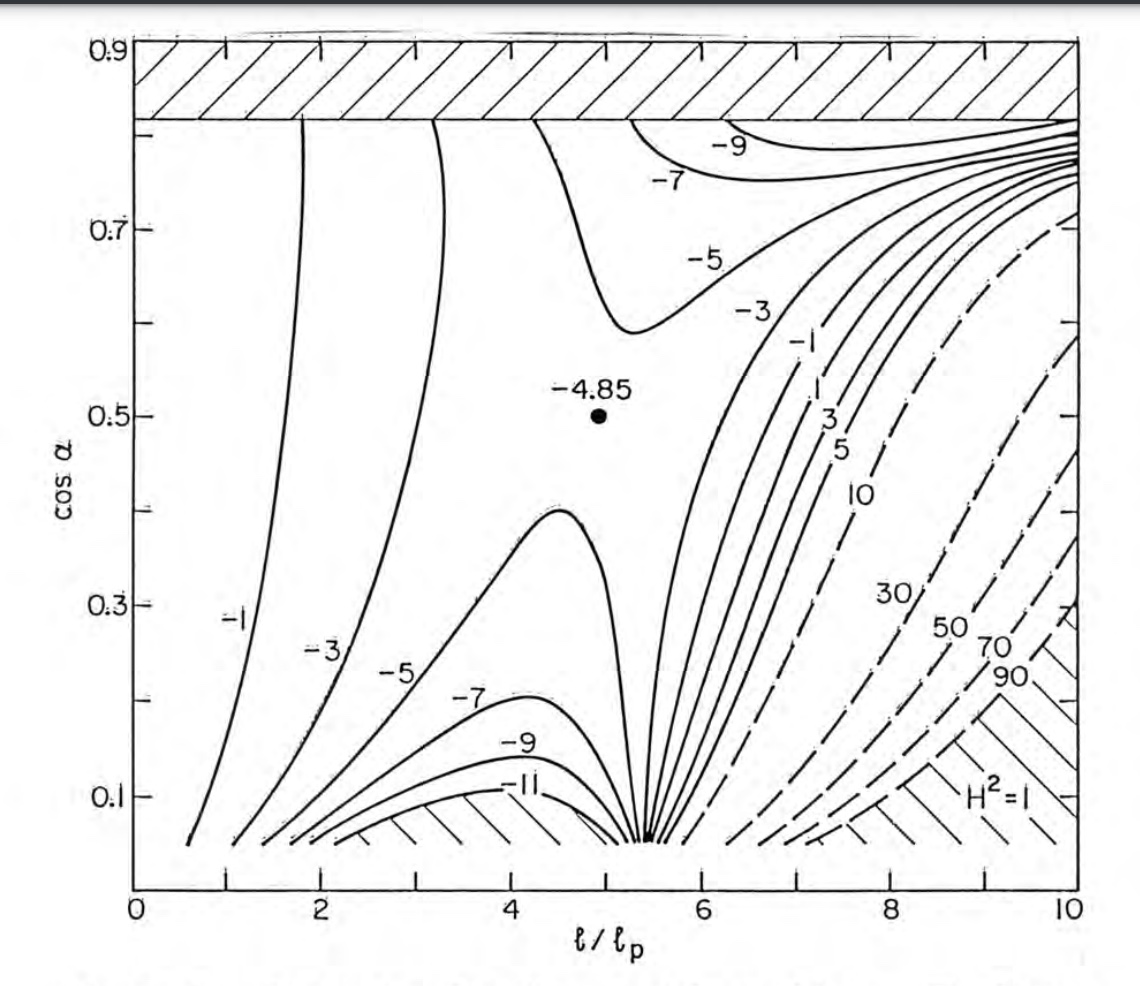}
  \caption{The action for distorted 5-simplices. The figure shows the action
	(divided by 100) for a two parameter family of 5-simplices in which all the
	edge lengths are $\ell$ except for the edges emerging from one vertex which
	are $\ell/(2\cos\alpha)$. $\alpha$ near $\pi/2$ corresponds to long thin
	5-simplices. $\alpha$ near 0 corresponds to nearly flat 5-simplices. There
	are no 5-simplices with $\cos\alpha$ greater than $.81$ because the 4-simplex
	inqualities would be violated. There is a saddle point corresponding to equal
	edges of value about 4.9. The edges at the saddle pint are thus  solution of the classical Regge equations  i.e. the simplicial Einstein equation. 
The negative gravitational action arising from conformal distortions is evident.
	\label{fig:3}}
\end{figure}

Figures \ref{fig:2} and \ref{fig:3} show numerical calculations of
Regge's action on the four-sphere. The simplest triangulations of $S^4$ are the
four dimensional surface of a 5-simplex $(\alpha_5)$ and the four dimensional
surface of the 5-cross polytope $(\beta_5)$---the 5 dimensional generalization
of the octohedron. These are the only regular solids in five dimensions. The 5
simplex has 6 vertices, 15 edges, 20 triangles, 15 tetrahedra and 6 4-simplices.
The cross polytope has 10 vertices, 40 edges, 80 triangles, 80 tetrahedra and 32
4-simplices. Figure~\ref{fig:2} shows the action for these triangulations as a
function of four volume when all their edges are equal and the cosmological
constant is unity in Planck units. The action is always lower than
the ``continuum" value corresponding to the round four sphere but becomes closer
to it as we move from the coarsest triangulation $\alpha_5$ to the finer $\beta_5$.

\eject
The edges  in Figure 3 have the value $\ell$ except those leading to a particular vertex which have the
value $\ell/(2\cos\alpha)$. $\cos\alpha$ near 0 thus corresponds to ``long and
thin" 5-simplices while large $\cos\alpha$ 5-simplices are ``short and squat."
$\cos\alpha$ cannot be too large because the analog of the triangle inequality
for 4-simplices would not be satisfied. The two parameter family shows the
characteristic saddle behavior of Einstein's action. There is an extremum when
all the edges are equal to about $4.84\ell_p$. This is a solution of the
discrete field equations corresponding to Euclidean de Sitter space. At this
solution the action is neither a maximum nor a minimum but a saddle point.
It is thus a solution of the discrete analog of the  Einstein eauation. 

One of the central features of geometric theories of gravity is their invariance
under the diffeomorphism group. In a simplicial approximation the diffeomorphism
group is broken in the sense that each different assignment of edge lengths will,
in general, correspond to a physically distinct geometries with distinguishable
curvatures. The diffeomorphism group reemerges in the limit of large $n_0$
because in this limit there are many simplicial geometries which {\it approximately}
correspond to a given continuum geometry and whose actions are {\it approoximately}
equal. The integrals in the numerator and denominator of (\ref{eq:6}) thus
approximately overcount continuum geometries in the same way that a sum over different
continuum metrics could overcount physically distinct geometries.

Let us see in more detail how this comes about. While in general one expects
different assignments of edge lengths to be different geometries, there is one
special case where this is certainly not true. This is flat space. Imagine
distributing vertices about a region of $n$-dimensional flat space, connecting
them so they form a simplicial manifold and assigning the appropriate flat space
distances between the vertices as edge lengths. If the vertices are now moved
about in flat space there will result a different assignment of edge lengths,
but this new assignment results in the same flat geometry. If
there are $n_0$ vertices in this part of the manifold there will be an $n_0$
parameter family of transformations of the edge lengths which leave the geometry
unchanged.

Consider a curved simplicial geometry with many vertices such that the typical
edge length is much smaller than the characteristic curvature scale $L$. For
example, in the process of solving the Regge equations on an increasingly
subdivided simplicial manifold to approach a continuum solution, one would
expect to reach such a geometry. (The Regge equation can, however, exhibit
solutions which do not correspond to a continuum one.   e.g.  Ref.(10). )
In this situation, regions small compared to the curvature scale will contain many
vertices and be {\it approximately} flat. There will therefore be
directions in the space of edge lengths in which  for each the action is {\it approximately}
constant for changes in the edge lengths smaller than the curvature scale.
These are the ``approximate diffeomorphisms" of the simplicial geometry. Their
number is correct --- $n$ directions for reach spacetime point.
In an expression like (\ref{eq:6}) we therefoere  expect each of the sums over
edge lengths in the numerator and denominator to diverge as $n_0$ becomes large. For
physical quantities, however, the ratio should remain finite.
\eject
\section{Summing over Topologies}

Summing over metrics is only one of two parts of a sum over geometries even as
the metric is only one of two parts in the specification of a geometry. The
other part might be loosely called the ``topology" and it is therefore of
interest to investigate sums over topologies. Simplicial approximation is a
natural framework in which to do this because the topological and metrical
aspects of a simplicial geometry are very clearly separated. The topological
information is contained in the rules by which the simplices are joined
together. The metrical information is contained in the assignment of edge
lengths. In particular, it is  possible to consider geometries with complicated
topologies but with relatively few edges.

To sum over the topologies of simplicial geometries with $n_0$ vertices is to
sum over some collection of simplices with a total of $n_0$ vertices. The widest
reasonable framework in which to discuss such collections is provided by the
connected simplicial complexes. A connected simplicial complex is a collection of
simplices such that if a simplex is in the collection then so are all its faces,
and such that any two vertices can be connected by a sequence of edges. What
connected complexes should be allowed? A natural restriction is to sum only over
complexes which are manifolds---that is, such that each point has a neighborhood
which is topologically equivalent (homeomorphic) to an open ball in $R^n$. In
classical general relativity, geometries on manifolds are the mathematical
implementation of the principle of equivalence. That principle tells us that
locally spacetime is indistinguishable from flat space, and this is the defining
characteristic of a manifold. It would, therefore, seem reasonable to consider
geometries on manifolds in the quantum regime although it is less clear that on
the scale of the Planck length the principle of equivalence should be enforced
in this strong way.

It is not straightforward to define a sum over manifolds. To do so there must at
least be an effective procedure for listing those manifolds which contribute to
the sum. We cannot do this by classification i.e. by taking ``one of type A",
``two of type B", etc.\/ because in four (and higher) dimensions the classification
problem for manifolds without additional structure is unsolvable. That is, there
does not exist an algorithm for deciding when two-manifolds are topologically
equivalent    (at least at the time  this was written.  Ref(5) ). This does not mean that one could not list topologies with more structure than manifolds for example conifolds Ref(13). Neither does it mean that one could not construct a list in 
which every manifold of a given dimension would be guaranteed to occur at least
once. One would simply not be able to tell when two manifolds on the list were
the same. Both of these approaches  have been suggested as possibilities for constructing
the sum over topologies. There is, however, another possibility: that we should sum
over a more general class of objects than manifolds.

In the sum over histories formulation of quantum mechanics we are familiar with
the idea of ``unruly histories." These are histories which contribute significantly
to the sums for quantum amplitudes but which are less regular than the classical
histories. For example, in particle quantum mechanics the dominant paths are
non-differentiable while the classical path is always differentiable. One would
perhaps be comfortable with admitting to a sum over topologies a larger class of
geometries than those defined on manifolds if one recovered manifolds in the
classical limit. The question is then: Is there a class of siimplicial  complexes
such that:
\begin{enumerate}
  \item the action for general relativity can be defined,
	\item there is an algorithm for listing the members of the class,
	\item manifolds are the dominant contribution to the sum over histories in
	the classical limit?
\end{enumerate}
I cannot yet answer this question in general. In two-dimensions, however, it is
easily addressed. This is because two-dimensional Einstein gravity has no metric
degrees of freedom. It is not, however, topologically trivial.

The Regge action extends naturally to any simplicial complex in two dimensions.
Recall that (Eq.(3) implies specically
\begin{equation}
   g_2I_2=-2\sum_{\sigma\epsilon\sum_0}\theta(\sigma)+2\Lambda\sum_{\tau\epsilon\sum_2}V_2(\tau)\label{eq:9}
\end{equation}
where the first sum is over the vertices and the second is over the triangles.
Insert the definition Eq(4)  in this expression, interchange orders in the
resulting double sum over vertices and triangles and note that the sum of the
interior angles of a triangle is $\pi$. One finds
\begin{equation}
   g_2I_2=-4\pi(n_0-n_2/2)+2\Lambda A,\label{eq:10}
\end{equation}
where $n_0$ is the number of vertices, $n_2$ the number of triangles and $A$ is
the total area. The curvature part of the action is independent of the edge
lengths and is therefore metrically trivial. The action, however, does depend
on how the simplices are joined together, that is, on the topology. This clean
separation of metric and topology makes two-dimensional Einstein gravity less
interesting than the higher dimensional cases but it also makes topological
questions easier to analyze.

\begin{figure}[htbp]
  \includegraphics[width=3.5in]{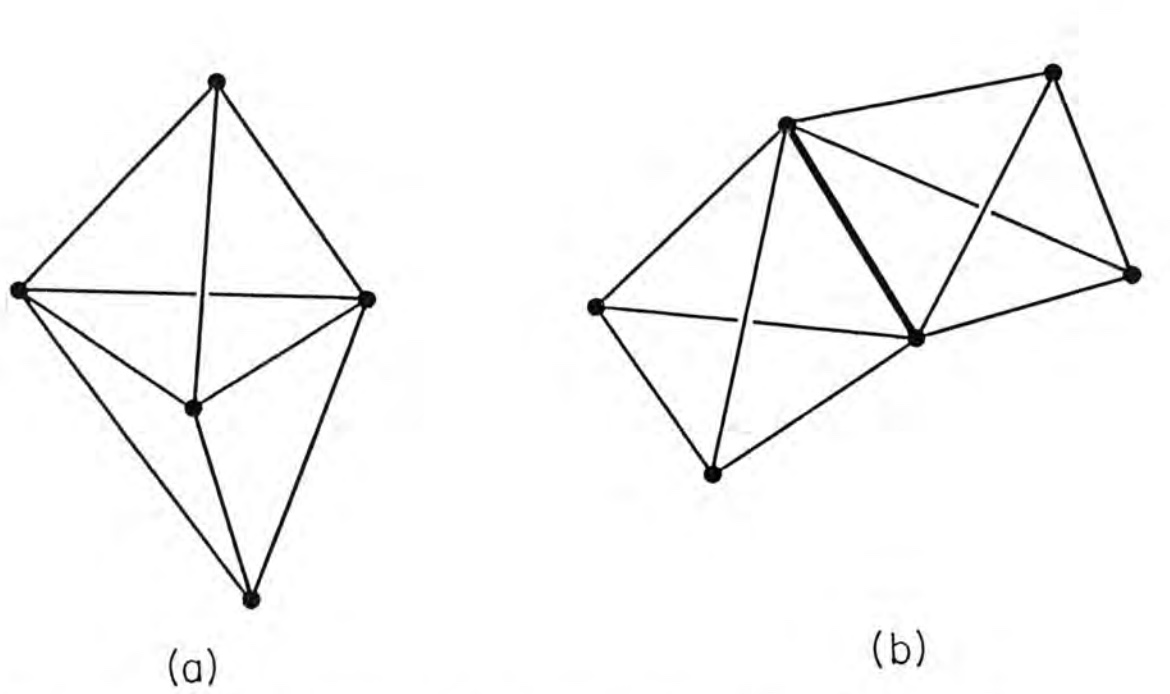}
  \caption{Branching and non-branching complexes. Non-branching two dimensional
	complexes like that in (a) have exactly two triangles intersecting at any one
	edge. The complex in (b) has four triangles intersecting along the more
	heavily drawn edge and is therefore a branching complex. Branching complexes
	fail to be manifolds at the edges on which they branch.
	\label{fig:4}}
\end{figure}

\begin{figure}[htbp]
  \includegraphics[height=3.5in]{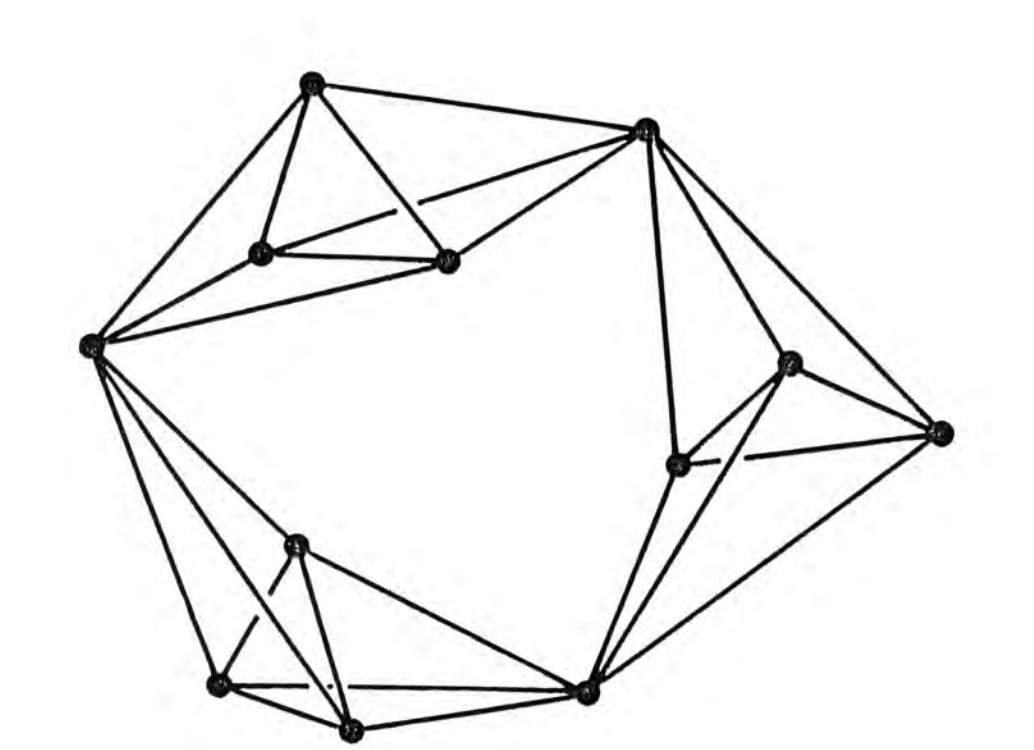}
  \caption{A two dimensional non-branching complex which fails to be a manifold
	at three vertices. This complex is not strongly connected and is thus not a
	pseudomanifold.
	\label{fig:5}}
\end{figure}

Let us start with simplicial complexes which are two-manifolds and enlarge the
class by giving up as little as possible until a larger class is found which
satisfies our criteria (\ref{eq:1}), (\ref{eq:2}), and (\ref{eq:3}) above. If a
complex is going to fail to be a manifold it must fail on some collection of
points. We give up least if we allow failure only at some discrete number of
vertices of the complex and do not permit failure along the edges. This means
we require every edge to be the face of exactly two triangles as in the complex
in Figure~\ref{fig:4}a. We thus exclude complexes like Figure~\ref{fig:4}b which
branch on an edge but permit those like Figure~\ref{fig:5} which fail at vertices.
For non-branching complexes, $3n_2=2n_1$ and the action is
\begin{equation}
   g_2I_2=-4\pi\chi+2\Lambda A\label{eq:11}
\end{equation}
where $\chi=n_0-n_1+n_2$ is the Euler number, a topological invariant.

\begin{figure}[htbp]
  \includegraphics[width=5.in]{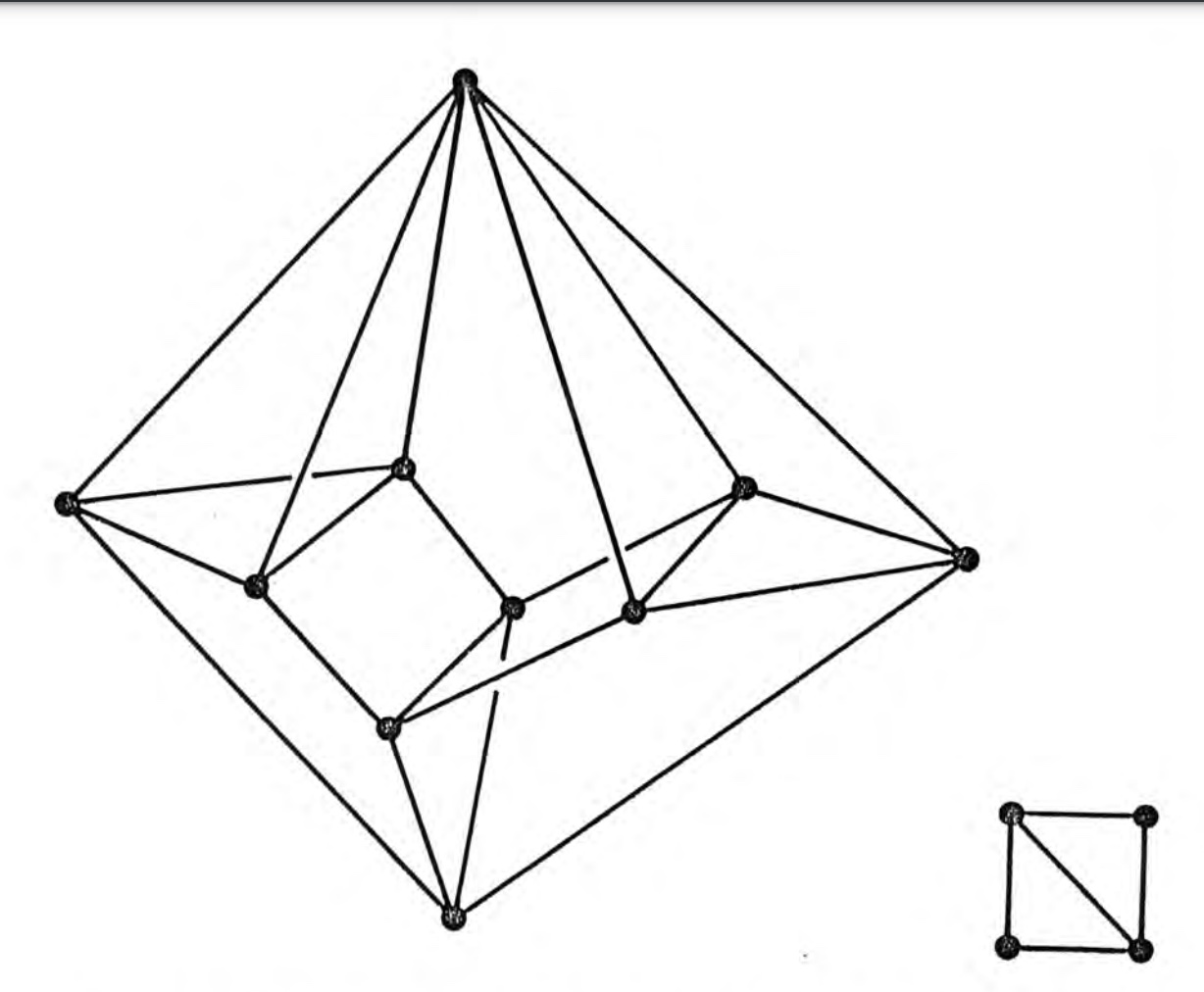}
  \caption{A pseudomanifold which fails to be a manifold at one vertex. The
	complex is two dimensional, non-branching and strongly connected. It is thus
	a pseudomanifold. It may be thought of as a sphere with two points identified.
         For pictorial clarity some of the edges triangulating
	quadrlaterals have been omitted but they should be imagined as in the example
	at lower right.
	\label{fig:6}}
\end{figure}

 Were we to stop here we could easily violate our criterion that a manifold
have the smallest action. Compare the sphere in Figure~\ref{fig:4}a which has
$\chi=2$, with the complex in Figure~\ref{fig:5}. It has $\chi=3$ and so a
smaller action.  it consists of almost disconnected pieces. To
prevent this we require that the complexes be strongly connected in the sense
that any pair of triangles can be joined by a sequence of triangles connected
along edges. The resulting complexes are called pseudomanifolds as described in  Ref.( 6).
The complex in Figure~\ref{fig:6} is a pseudomanifold whereas the one in Fig 5 is not.

In two-dimensions, pseudomanifolds have $\chi\ge2$,
and the pseudomanifold with $\chi=2$ is the sphere. Thus the pseudomanifold
with the smallest action is a manifold and we recover manifolds in the
classical limit.

Most importantly for us, however, pseudomanifolds are easily enumerable. Their
defining properties in $n$ dimensions are

\begin{enumerate}
  \item Pure dimension---a simplex of dimension $k<n$ is contained in some $n$-simplex.
	\item Nonbranching---an $(n-1)$-simplex is the face of exactly two $n$-simplices.
	\item {Strongly connected---any two $n$-simplices can be connected by a sequence \\ of $n$-simplices connected along $(n-1)$-simplices.}
\end{enumerate}

These defining properties are essentially combinatorial. Given $n_0$ vertices
one can imagine listing all the possible collections of $n$-simplices and checking
to see which are pseudomanifolds and which are not in a finite number of steps.

In two-dimensions pseudomanifolds satisfy all three criteria for a class of
complexes with which to define a sum over topologies. The Regge action is
defined for them, there is an algorithm for enumerating them, and the pseudomanifold
of least action is a manifold. In higher dimensions, finding a class which meets
these criteria is a deeper question. Pseudomanifolds can be defined in higher
dimensions as described above. The action can be extended to them and they are
enumerable. Finding the configurations of least action, however, is now not only
a question of topology but also of metric. The possibilities for pseudomanifolds
are so varied in higher dimensions that it seems likely that one must restrict
the class of complexes further in order to have manifolds dominate in the classical
limit. Then by relaxing the principle of equivalence at the
quantum level we will have an attractive class of geometries with which to define
a sum over topologies in quantum gravity.

\section*{Acknowledgments}
Thanks to Debbie Ceder who transformed the original into latex. Preparation of this report was supported in part by the National Science Foundation, under grant PHY 81-07384. The author is indebted to  Ruth Williams for  conversations on the Regge Calculus over many years and for a critical reading of this paper.


\end{document}